# Juxtaposition of Shallow Reservoir-Triggered Seismicity and Deep Tectonic Locking in the Qiaojia-Dongchuan Seismic Gap

Yuxin Zhou [1,2], Huai Zhang [1*], S. Mostafa Mousavi [2], Guangyao Yin [1], Pei He [1], Yicun Guo [1], Shuang Yi [1], and Yaolin Shi [1]

[1] State Key Laboratory of Earth System Numerical Modeling and Application, College of Earth and Planetary Sciences, University of Chinese Academy of Sciences, Beijing, China 100049.

[2] Department of Earth and Planetary Sciences, Harvard University, Cambridge, MA 02128, USA

[*] *Corresponding author: Huai Zhang (hzhang@ucas.ac.cn)*

**Abstract**

Identifying the critical state of mature seismic gaps is challenging, especially when anthropogenic stress perturbations, such as reservoir impoundment, superimpose on tectonic loading. Here, utilizing a high-resolution dense array catalog from the Qiaojia-Dongchuan seismic gap (hosting the world's second-largest hydropower station), we reveal a distinct vertical decoupling mechanism. The shallow activities exhibit high b-values (1.0), indicative of fluid-driven reservoir-triggered seismicity. Conversely, deep seismicity (15–20 km) outlines a "locked asperity" characterized by low b-values (<0.8) and high Coulomb stress accumulation rate (~10 kPa/yr). We further identify a complex dipping structure, suggesting compound fault kinematics. Additionally, the calculated stress accumulation suggests this seismic gap is in a critical state with elevated rupture potential. Our findings indicate that shallow induced seismicity can mask the silent accumulation of deep tectonic strain. This decoupling model provides a new framework for assessing seismic risks in reservoir-fault systems globally.

## Introduction

The global expansion of large-scale hydroelectric projects into tectonically active regions poses a fundamental challenge for seismic hazard assessment. Especially when massive reservoirs are impounded over mature seismic gaps[1]—fault segments that have accumulated significant strain but remained quiescent—the interaction between anthropogenic stress perturbations and deep tectonic loading is highly complex. While Reservoir-Triggered Seismicity (RTS)[2,3] is a well-documented phenomenon characterized by fluid diffusion and poroelastic loading, its interaction with a critically stressed, locked fault remains poorly understood. A critical unresolved question is whether reservoir impoundment triggers the premature release of accumulated tectonic strain (potentially leading to large earthquakes) or merely activates shallow secondary fractures, thereby masking the silent accumulation of deep tectonic energy. Distinguishing between these two scenarios is vital, as the latter implies a hidden, growing risk that standard monitoring might misinterpret as stress release.

The Qiaojia-Dongchuan segment of the Xiaojiang Fault represents a rare and critical geological "laboratory". First of all, it is identified as a mature seismic gap[4–6] since it has been historically active but quiescent in recent decades (Fig. 2C), accumulating significant tectonic strain. The recent Mw 7.7 earthquake that struck northern Myanmar in March 2025 occurred within a long-recognized seismic gap[7,8]. This event, which resulted in significant casualties, heightened interest in seismic gap research. The Qiaojia-Dongchuan region represents a critical tectonic gap within the eastern margin of the Tibetan Plateau and the Sichuan Basin. From 1970–2000, the Sichuan–Yunnan block (Fig. 1), as one of China's most seismically active areas[9–11], experienced thousands of M≥5 events and 36 clustered M≥7 events, but only 5 M≥7 earthquakes occurred between 2000 and 2023[12,13]. The northern Longmenshan fault experienced the 2008 Ms 8.0 Wenchuan (focal depth ≈14–19 km) [14,15] and 2013 Ms 7.0 Lushan earthquakes (focal depth ≈13–17 km) [9]. However, some segments remain seismically quiescent, including areas between the Longmenshan and Lijiang-Xiaojinhe fault zones, the middle–southern Xianshuihe fault zone in Fig. 1A, and the Shimian–Xichang section (Gap 1 in Fig. 1A) of the Anninghe fault zone. The 2022 Mw 6.6 Luding earthquake, which occurred at a depth of approximately 12–16 km, ended a 236-year quiescence on the southern Xianshuihe fault[16], but the Xiaojiang fault remains unruptured.

Second, characterized by its location at the intersection of 5 major faults (Fig. 2), which includes several major NNW to SSE striking faults such as the Daliangshan-Jiaojihe fault, Anninghe-Zemuhe fault, Puduhe fault, and Xiaojiang faults[17–19], the Qiaojia-Dongchuan seismic gap is accumulating significant tectonic strain. Historically, the Zemuhe-Xiaojiang fault system has been associated with significant seismic events, recording four earthquakes of M ≥ 7.0 between 1733 and 1850. These include the 1733 M7.8 Dongchuan earthquake[20], the 1789 M7.0 Huaning earthquake, the 1833 M8.0 Songming earthquake[20,21], and the 1850 M7.5 Xichang earthquake[22]. Since then, only moderate earthquakes have occurred, notably the 1909 M6.7 Huaning and the 1966 M6.5 Dongchuan events[4]. These events largely exhausted the seismogenic segments of the fault system, which then entered a long-term interseismic strain-accumulation phase [5]. Actually, previous studies[16,23,24] have shown that the Qiaojia-Dongchuan and northern Jianshui segments along the Xiaojiang fault exhibit intense surface deformation (measured via GNSS) consistent with significant tectonic loading rather than fault locking[17]. These regions show clear signs of mechanical locking and significant strain deficits, indicating the potential for future large earthquakes (M ≥ 7.0). However, most research has focused on the seismic gap along the Daofu–Shimian segment of the Xianshuihe fault and the northern Jianshui segment [25,26], but the Qiaojia-Dongchuan region remained largely unexplored.

Most importantly, this gap hosts the newly commissioned Baihetan Hydropower Station—the world’s second largest—providing a unique opportunity to observe the immediate interaction between extreme hydraulic loading and a locked fault system. The Baihetan Dam is a massive double-curvature arch dam (the total capacity of over 20 billion cubic meters) on the Jinsha River in southwest China, and its associated hydropower station is the second largest in the world by installed capacity. It became fully operational on December 20, 2022. Reservoir-Triggered Seismicity (RTS) has long been a subject of concern since the 1960s, following devastating earthquakes exceeding magnitude 6.0 at Kremasta (Greece)[27] and Koyna (India)[28–30], the latter of which resulted in approximately 200 fatalities and significant infrastructure damage. However, previous studies have examined RTS in stabilized regions or specific events[30,31]. Our study captures the immediate seismic response of a "locked" tectonic asperity to rapid impoundment. The accumulation of tectonic strain to near-critical levels make seismic gap zones disproportionately sensitive to the anthropogenic stress perturbations introduced by large reservoirs. Reservoir impoundment imposes additional water load and drives pore-pressure diffusion that can, by itself, erode the residual frictional strength of a locked fault; when the fault occupies such a structural gap, this effect is further amplified, potentially triggering seismic rupture significantly earlier than natural tectonic processes would permit[32].

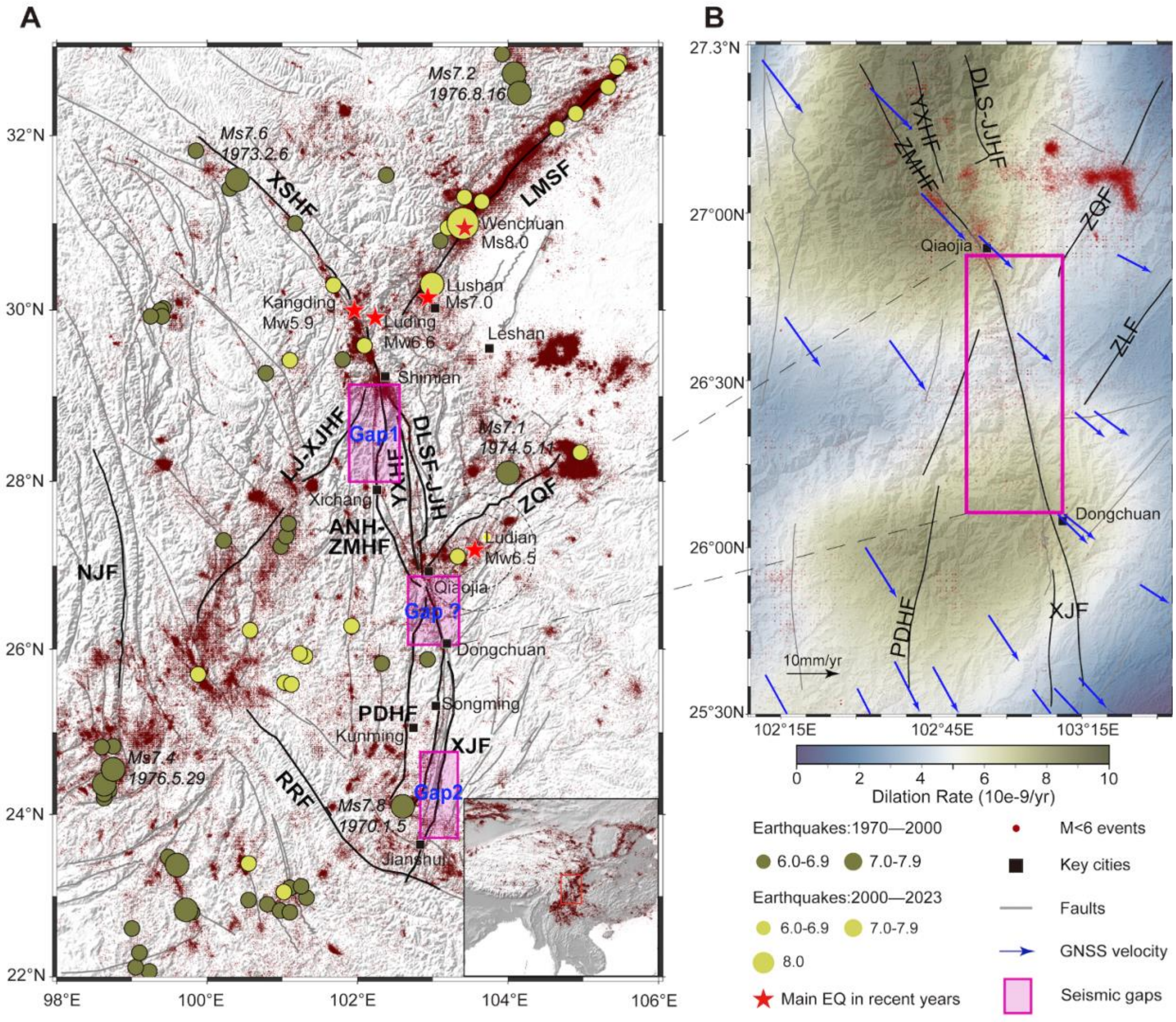

**Fig. 1.** Distribution of seismic events and tectonic loading map in the study area. (**A**) The complex active fault system with the latest fault data from Lu *et al.*[33]. XSHF: Xianshuihe Fault; LMSF: Longmenshan Fault; LJ-XJHF: Lijiang-Xiaojinhe Fault; ANH-ZMHF: Anninghe-Zemuhe Fault; YXHF: Yuexihe-Heihe Fault; DLSF-JJH: Daliangshan Fault-Jiaojihe segment; NJF: Nujiang Fault; ZQF: Zhaoqiao Fault; ZLF: Zhaotong-Ludian Fault; PDHF: Puduhe Fault; XJF: Xiaojiang fault; RRF: Red River fault. The earthquake catalog (hereafter referred to as the standard catalog) derived from the National Earthquake Data Center of the China Earthquake Administration (http://10.5.160.18/console/exit.action). The regional context map indicates the study area (red box) within the broader tectonic framework of Southeast Asia. (**B**) Tectonic loading map

of the Qiaojia–Dongchuan seismic gap region and surrounding areas in the study. Blue arrows indicate GNSS-derived horizontal velocity vectors indicating ongoing crustal deformation, based on data from Wang *et al.*[34] and Zhang *et al.*[35]. The background color corresponds to the dilatation rate ($10^{-9}/year$), with positive values (green) indicating compressive deformation (data from Zhang *et al.*[35]).

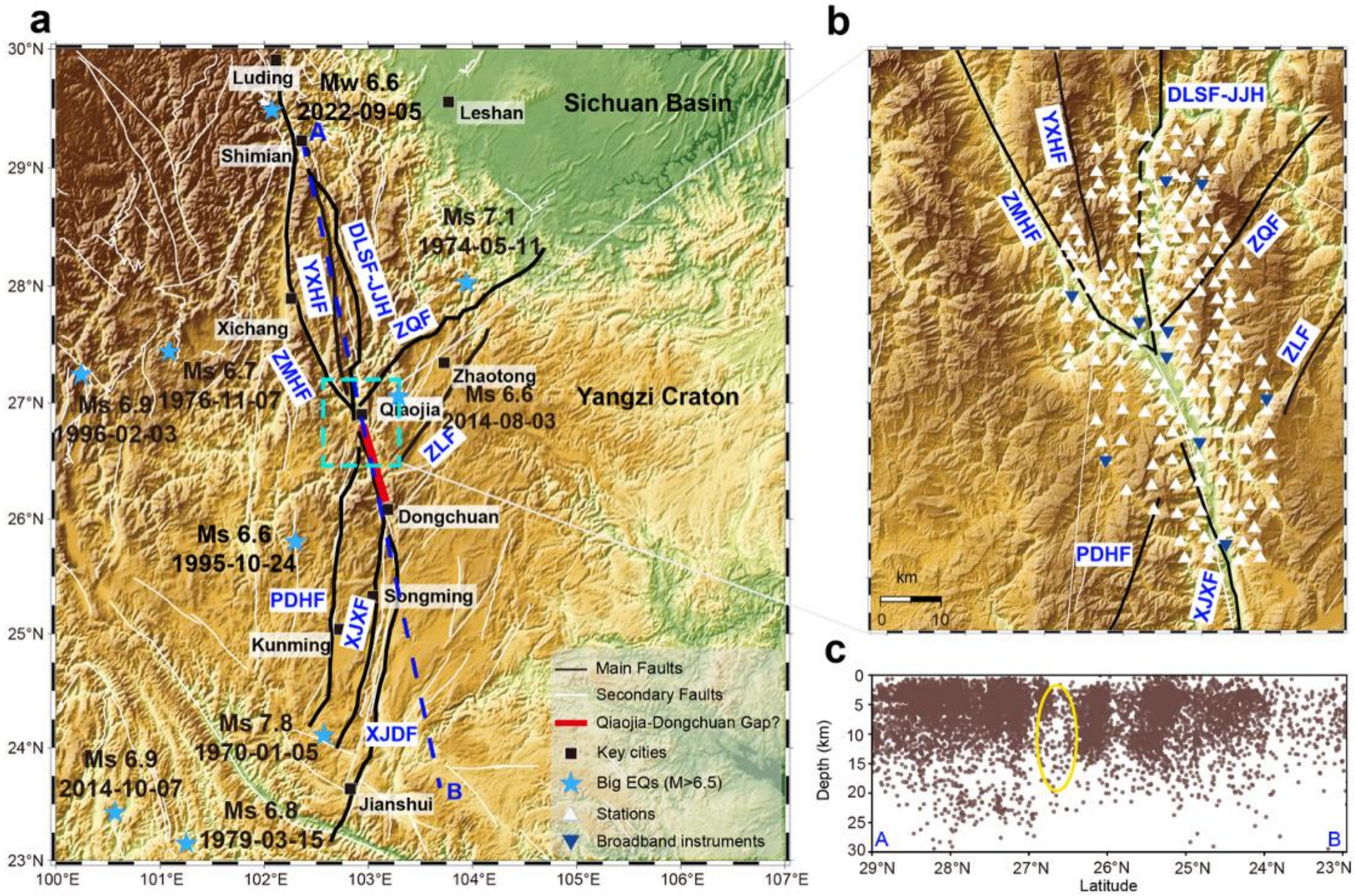

**Fig. 2.** Tectonic features and the distribution of the deployed Broadband stations with short-period dense array in the study area. (**A**) The topographic map of the study area. Filled black squares represent major cities, with solid white lines indicating secondary faults in the vicinity of the Qiaojia-Dongchuan region. Black solid lines represent major faults. ZMHF: Zemuhe Fault; YXHF: Yuexihe-Heihe Fault; DLSF-JJH: Daliangshan Fault-Jiaojihe segment; ZQF: Zhaoqiao Fault; ZLF: Zhaotong-Ludian Fault; PDHF: Puduhe Fault; XJXF: Xiaojiang West Fault; XJDF: Xiaojiang East Fault. The red line segment indicates the Qiaojia-Dongchuan seismic gap. (**B**) Distribution of seismic stations in the Qiaojia-Dongchuan region. White triangles represent the locations of short-period stations, and blue inverted triangles represent the locations of the broadband instruments used in this study. (**C**) Cross-sectional distribution of the standard catalog from 1970 to 2023 along depth profile A–B. The yellow ellipse represents the gap in events distribution at Qiaojia-Dongchuan segment.

Given this context, we utilize the Qiaojia-Dongchuan segment as a unique natural laboratory. To capture the mature seismic gap's response of the complex fault system to this reservoir project, we leveraged a high-resolution catalog derived from a dense array of over 200 seismometers deployed during the peak impoundment phase (August 2022–March 2023) (Fig. 2B). By integrating this monitoring with Coulomb stress evolution modeling and detailed spatiotemporal Gutenberg–Richter b-value[36] analysis, we reveal a distinct vertical decoupling mechanism. Furthermore, we image a previously unrecognized dipping structure, suggesting compound fault kinematics. Our findings demonstrate that shallow induced seismicity can effectively mask deep tectonic locking, providing a new conceptual framework for assessing seismic risks in reservoir-fault systems globally.

## Results

### The high-resolution microseismic catalog

After processing the data through the ensemble machine learning workflow (see materials and methods) and applying strict threshold filtering, 5,219 seismic events were detected and precisely located in the study area between August 2022 and March 2023 (Fig. 3). This is almost an order of magnitude more events than the existing events (532), for the same period, in the seismicity catalog of Chinese earthquake data center (hereafter referred to as "standard catalog"). The AI-based approach significantly improved the detection capability, particularly for microseismic events ($M_L$ < 3.0). The seismicity catalog generated by the AI method exhibits a high degree of spatial consistency with the standard catalog, both in the overall distribution (Fig. S1) and in the depth distribution (Fig. 3D), while significantly expanding the total number of recorded events. The overall magnitude of completeness (Mc) calculated by max curvature is ML 1.29, and the overall b-value of the catalog is 0.86+/-0.01(Fig. 3B). The spatial variations of Mc are shown in Fig. S2.

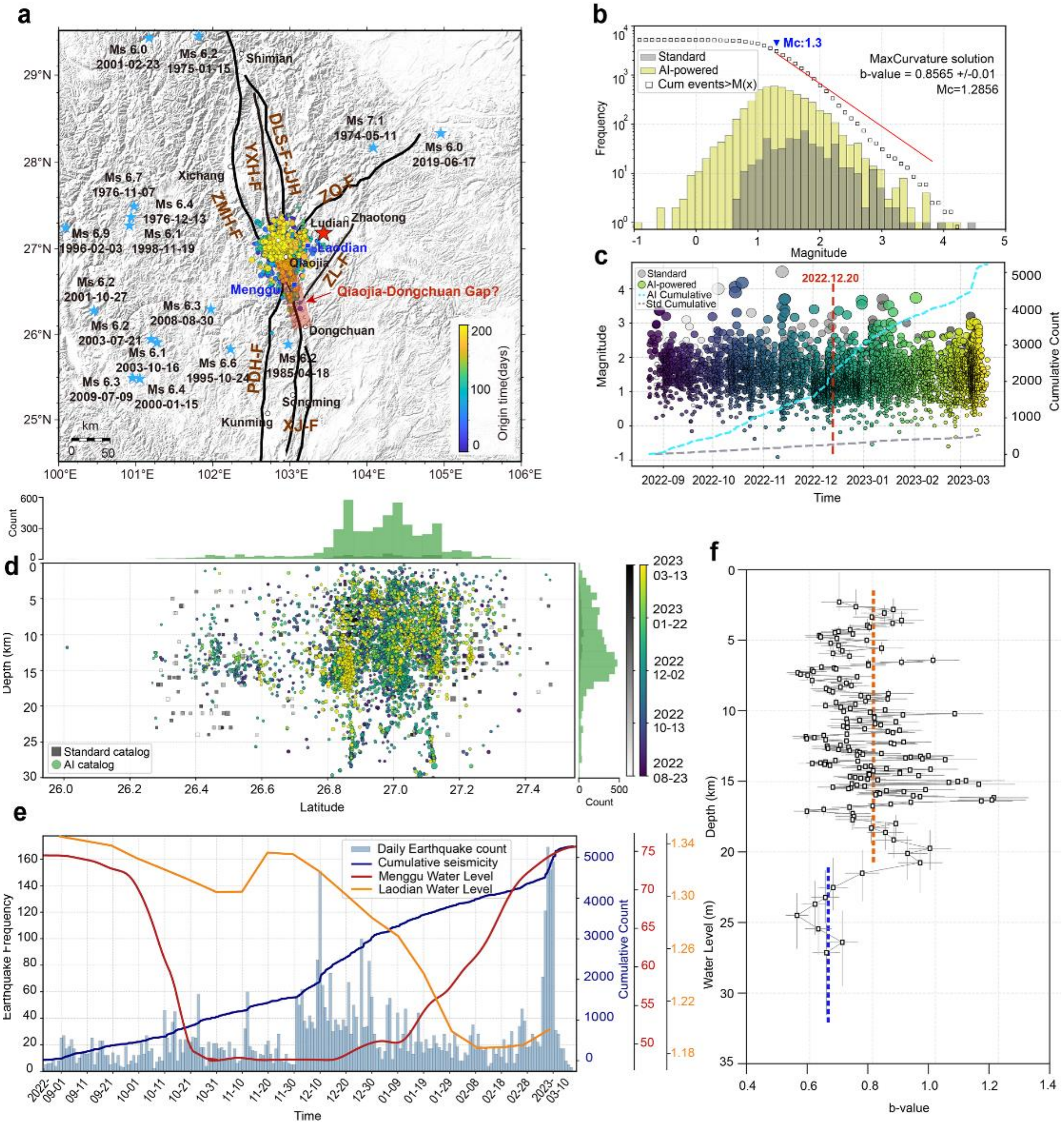

**Fig. 3.** AI-detected earthquake event catalog and distribution. (**A**) Locations of events following rigorous absolute and relative relocations. The events are color-coded by origin time. The marker size is scaled by magnitude, with black lines representing faults: ZMHF: Zemuhe Fault; YXHF: Yuexihe-Heihe Fault; DLS-F-JJH: Daliangshan Fault-Jiaojihe segment; ZQF: Zhaoqiao Fault; ZLF: Zhaotong-Ludian Fault; PDHF – Puduhe Fault; XJF – Xiaojiang Fault. The light blue pentagrams denote the major historical earthquakes with magnitude ≥6.0. The dark blue square indicates the Menggu wharf and Laodian wharf near Qiaojia. (**B**) Magnitude distribution of the detected events in this study, compared with the events in the national seismic catalog, and the frequency-magnitude distribution of the detected events in the catalog. "Standard"

refers to the catalog from the National Earthquake Data Center (https://data.earthquake.cn/), while "AI-powered" denotes the catalog generated in this study. (**C**) Magnitude-Time plot of two catalogs. The marker size is coded by magnitude, while the color codes by origin time. 'Std cumulative' in the legend denotes the cumulative curve corresponding to the standard catalog. (**D**) The depth profile of the standard catalog and the AI catalog. The gray squares indicate the standard catalog, and the green circles indicate the AI catalog. (**E**) Earthquake frequency and water level change. The red line and the orange line indicate the water level changes at Menggu wharf and Qiaojia Laodian wharf separately. (**F**) The b-value as a function of depth, calculated by ZMAP[37].

Fig. 3C shows a sharp peak in seismic activity during December 2022, coinciding with the completion of the Baihetan power station (on December 20, 2022), indicating a potential rapid elastic response of the fault systems. A secondary clear spike in seismic activity is evident in March 2023, potentially related to delayed pore-pressure diffusion due to the reservoir impoundment. However, this temporal correlation alone does not constitute definitive proof of a diffusive mechanism, and alternative explanations cannot be excluded (see Discussion). Regardless of direct correlation with water level and reservoir construction, this temporal behavior of the seismicity suggests new stress perturbations or changes in the regional tectonic stress. The seismic activity predominantly consists of shallow-focus events, with about 70% occurring at depths of less than 15 km. More events are observed in the shallow layers (0-20 km), possibly suggesting a near-surface source of this stress perturbation, concentrating in the overlying strata and in fault activity, whereas in the intermediate to deep layers (20-30 km), events are sparse. The distribution of seismic events on the horizontal plane displays a distinct non-homogeneous pattern characterized by localized high-density zones. In addition, many newly detected microearthquakes are densely clustered around latitude 26.9° — corresponding to the Qiaojia area — and exhibit an apparent vertical alignment (Fig. 3D).

## Spatiotemporal Distribution of the Seismicity

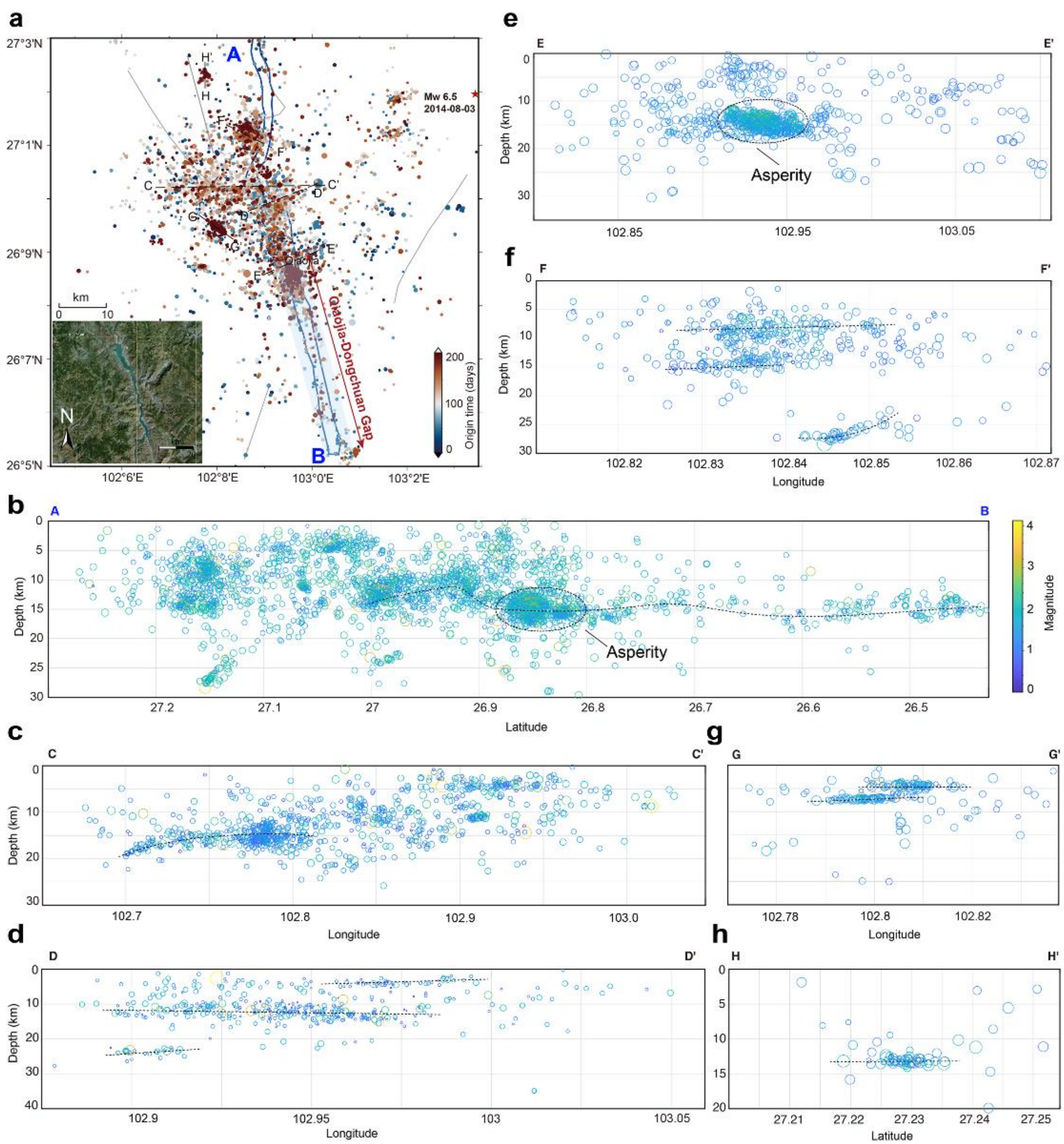

**Fig. 4.** Distribution of earthquake event locations and depth profiles in AI catalog. (**A**) The spatial distribution of earthquake events in AI catalog along potential profiles. The event color is determined by the origin time. The lower-left corner shows the satellite image of the study area, with the blue line indicating the outline of Baihetan Reservoir and the red star marking the position of the 2014 Ludian Mw 6.5

earthquake. (**B**) Cross sections along the outline of the reservoir A-B from north to south. (**C**-**H**) West-east and north-south cross sections marked in the map.

The high-resolution earthquake catalog (Fig. 4) reveals a distinctly heterogeneous pattern of seismicity across the study area. Seismic events are intensely concentrated along a major NNW–SSE-trending structural corridor, primarily controlled by the Xiaojiang Fault Zone. The outline of the Baihetan Reservoir (blue solid line in Fig. 4A) aligns remarkably well with the trace and orientation of this fault zone, suggesting a close spatial correspondence between the reservoir and the primary tectonic feature. The Xiaojiang Fault extends roughly 20 km along strike, exhibiting a pronounced NNW orientation, and represents the most dominant tectonic structure in the region. The Xiaojiang Fault Zone shows superior continuity in both depth distribution and along-strike clustering of seismicity (Fig. 4B). Specifically, a continuous seismicity belt is observed at depths of 15–20 km. Previous studies[38] have also delineated the depth extent of the Xiaojiang Fault Zone as 10–25 km based on the depth distribution of micro-seismic events. We therefore infer that this belt represents the along-depth outline of the Xiaojiang Fault Zone. Although several secondary faults (gray lines in Fig. 4A) appear to form a near-conjugate configuration, the Xiaojiang Fault should be regarded as the principal active structure rather than merely part of a conjugate system.

Depth profiles (Fig. 4B–H) show that most earthquakes occur between 5 and 15 km, forming a well-defined rupture zone with high spatial coherence. Multiple cross-sections (C–C', F–F') reveal a persistent seismic belt with arc-shaped clustering, and profiles such as E–E' and A–B' highlight a compact asperity—a zone of concentrated microseismicity surrounded by quiescence—at approximately 12–18 km depth suggesting localized stress accumulation and sufficient strain loading. Notably, the Qiaojia-Dongchuan seismic gap is identified near the northern and western segments of the Xiaojiang Fault (Fig. 4A). Despite relatively active seismicity on both sides of this zone, the interior shows a persistent lack of moderate-to-strong earthquakes, forming a spatial pattern characteristic of stress concentration zone. The consistency of this feature across multiple profiles implies that the gap may host elevated strain energy and could therefore represent a potential locus for future larger earthquakes.

**Coulomb Stress Evolution and B-value Variation of the Qiaojia–Dongchuan Seismic Gap**

The calculated annual Coulomb stress rate (Fig. 5A) reveals a broad positive CFS rate zone centered south of Qiaojia, extending southwestward toward Dongchuan, with values exceeding 8 kPa/yr. The highest rates (~10 kPa/yr) are concentrated along the central section of the Zemuhe–Xiaojiang fault, specifically in the northern part of the Qiaojia–Dongchuan segment. Moderate positive CFS rate values of 3–6 kPa/yr are observed westward along the Zemuhe Fault and northward along the Jiaojihe Fault. In contrast, a localized negative CFS rate anomaly is identified in the northeastern portion of the main fault segment, east of the Zhaoqiao Fault. The transition from positive to negative values is particularly sharp in the northeast quadrant, whereas the western sector maintains uniformly positive values over a broad area. This pattern highlights a persistent mechanism of stress accumulation in the central and southwestern parts of the fault system—specifically localized at the fault junction and the seismic gap—with stress release confined to a narrow zone in the northeast.

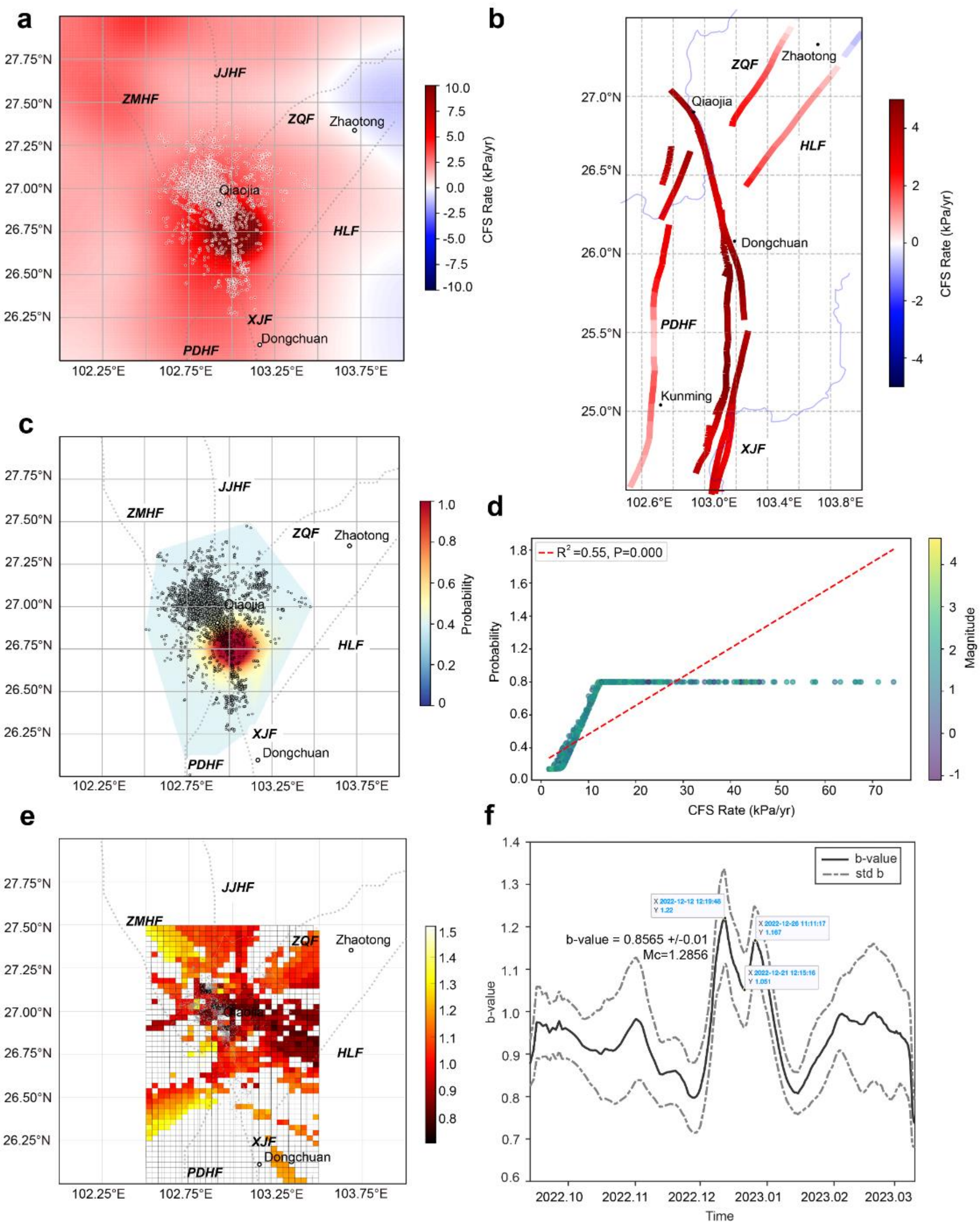

**Fig. 5.** Coulomb stress rate, earthquake triggering probability, and statistical validation in the Qiaojia–Dongchuan segment. (A) Spatial distribution of Coulomb Failure Stress Rate (CFS rate). Positive values

(red) indicate stress accumulation conducive to failure. (B) CFS rates values along individual fault segments of the Xiaojiang Fault Zone. (C) Earthquake triggering probability distribution. (D) Scatter plot showing the relationship between CFS rates and triggering probability at earthquake locations. The fitted red dashed line indicates a significant positive correlation ($R^2$ = 0.55, $p$ < 0.001). Marker color denotes earthquake magnitude. (E) The b-value distribution map using ZMAP software[37]. (F) The b-value as a function of time calculated using ZMAP. Dashed lines show standard deviations.

We then statistically validated the constructed earthquake catalog against the computed CFS rate field (Fig. 5A). All events were located within areas of positive CFS rates, with 93.7% of events situated in areas of high CFS rates. This independent result not only supports the spatial reliability of the detected earthquakes but also strengthens the case for using CFS rates as a quantitative metric for seismic hazard assessment for southwestern China. The high CFS rate area collocates with a low b-value range (0.8–0.95). However, no abnormal b-value is observed in the localized area with peak CFS rates, nor is the probability (Fig. 5E). Fig. 5B shows that most segments of the Xiaojiang Fault exhibit positive CFS rates, indicating a stress-loading regime. Notably, the area south of Qiaojia exhibits sustained stress accumulation, with local CFS rates reaching 7–10 kPa/yr, consistent with previous studies[39,40].

The resulting triggering probabilities depend on the choice of parameter settings, particularly the value of direct effect parameter $A_\sigma$ and the recurrence interval $\tau$. Fig. 5C shows the triggering probabilities obtained with $A_\sigma = 0.3 \times 10^6$ and $\tau = 10$. Under this parameter configuration, the triggering probabilities range from 0.4007 to 1.0, with a mean value of 0.5107. Owing to parameter uncertainty, the distribution of mean triggering probabilities across different parameter settings is shown in Fig. S3. A substantial proportion of events occur in areas with positive CFS rates, and statistical significance analysis ($R^2 = 0.55$, $p$ < 0.001) reveals a strong correlation between seismicity and CFS rate evolution. Temporal variation of b-value shows two distinct peaks, corresponding to two surges of seismic activities (Fig. 5F). The separation of estimated temporal b-value at the beginning and the end of the observed seismicity indicates the high uncertainty due to the short duration of the catalog and/or variations of the magnitude of completeness as a function of time or space.

## Discussion

### Potential Induced (Reservoir-Triggered) Seismicity at Baihetan Reservoir

Construction of the Baihetan Hydroelectric Power Station began in 2010, with the main phase initiated in 2017. The first batch of units became fully operational on June 28, 2021, and all units were officially commissioned on December 20, 2022. The reservoir impoundment phase commenced in July 2022, with the main impoundment occurring in January 2023. Our seismic catalog shows consistent activity between September and November 2022, concurrent with the active impoundment phase of the reservoir (Fig. 3E). However, a distinct change occurred in December 2022, coinciding with the peak of reservoir impoundment and the station's full commissioning. During this period, seismicity frequency reached its highest level, exceeding 300 events per month in the AI catalog. This seismic peak was characterized by a "high-frequency, low-magnitude" pattern. The magnitudes did not significantly increase (ML < 3.5), and the proportion of small-magnitude events (ML < 2.0) was notably high. It is quantitatively captured by a peak b-value of 1.2 (Fig. 5F). Spatially, the activity formed a concentrated cluster around 26.9°N—corresponding directly to the main dam area—with events primarily located at shallow and mid-shallow depths near the reservoir edge (Fig. 4A).

The spatial alignment of seismicity with the Baihetan Reservoir and the temporal correlation with its impoundment provide circumstantial evidence suggestive for RIS. RIS typically operates through two distinct physical processes: a) poroelastic (elastic) loading, an immediate response to the weight of the water column that instantly alters the crustal stress tensor; and b) pore-pressure diffusion, a delayed response caused by the migration of fluids into fault zones, reducing effective normal stress, which usually manifests months to years after impoundment[32,41].The rapid response observed in December 2022 is consistent with an elastic loading mechanism or rapid pressure transmission through highly permeable fracture networks. The secondary peak in March 2023, occurring approximately 3–4 months after the peak impoundment, is temporally consistent with a delayed pore-pressure diffusion response. Analogous delayed responses have been documented at other reservoirs: for instance, a ~4.5-month delay at the Açu Reservoir in Brazil[42] and a ~75-day delay at the Nurek Reservoir in Tajikista[43]. However, we note that a formal verification through $r \sim \sqrt{t}$ analysis has not been performed for the Baihetan dataset, and this remains a priority for future investigation. The high b-value (1.2) and swarm-like behavior are characteristic of the fluid-driven activation of shallow fractures or local stress-field adjustments within active tectonic regions[44,45]. Furthermore, numerical modeling by Lui *et al.*[46] indicates that pore-pressure disturbances during seismic quiet periods can lead to both early and delayed subsequent seismic events. Similar characteristics have been documented elsewhere; a previous review[3] summarized the evolution of seismicity at the Hoover Dam, noting a link between early impoundment (1935–1945), shallow depths (<5 km), and pore-pressure diffusion. The concentration of shallow seismicity at Baihetan aligns with these established RIS characteristics[47,48].

Distinguishing the mechanism of this seismicity is crucial for hazard forecasting. It is critical to differentiate between "induced" events (where reservoir stress change is the primary cause) and "triggered" events (where the reservoir provides the final perturbation to a critically stressed tectonic fault). The decreasing b-value trend in depth may imply spatial decoupling: a shallow reservoir zone characterized by high b-values and fluid-driven microswarms (stress release), versus a deep tectonic fault (seismic gap) characterized by low b-values and tectonic loading (stress accumulation) (Fig. 3F).

A linear seismic cluster near the Xiaojiang Fault (XJF) suggests some events are distributed along the fault, potentially reflecting the combined effects of tectonic stress and reservoir loading. The December 2022 peak consisted of small magnitudes ($M_L < 3.5$) that did not release significant moments, supporting the interpretation that the reservoir is activating secondary structures (likely dipping faults) rather than releasing the primary moment deficit of the XJF. Regarding the water volume, the Baihetan Reservoir has a total capacity of over 20 billion cubic meters with a normal water level of 825 m [49]. Given its large surface area, even modest water level fluctuations (on the order of a few meters) correspond to substantial volumetric changes capable of generating significant poroelastic loading. Consequently, the reservoir may act as a "stress probe," illuminating the crust's critical state without relieving accumulated tectonic strain. Coulomb stress modeling confirms that impoundment increases CFS rate on the deep

Xiaojiang fault (Fig. 5A). This suggests the reservoir load may be acting to "clamp" or load the deeper Qiaojia-Dongchuan segment—potentially bringing the locked asperity closer to failure—while simultaneously inducing shallow seismicity.

Previous studies on RIS in China have primarily focused on other large reservoirs, such as the Three Gorges and Gezhouba reservoirs on the Jinsha River[50–52]. However, triggering mechanisms at Baihetan may exhibit unique characteristics due to the specific tectonic settings and local geological conditions of the Qiaojia-Dongchuan region. While the clustering of seismicity provides preliminary evidence for RIS, a more integrated analysis is necessary. As noted by Yi et al.[53], RIS is related to changes in load and pore pressure but is also influenced by the regional tectonic background. Local weak zones in the geological structure may intensify the impact of RIS, resulting in the observed elevated seismic activity.

**Fault Structure**

The cross-sectional profiles (Fig. 4B–F) provide new insights into the subsurface fault geometry that may play a critical role in controlling strain accumulation within the Qiaojia–Dongchuan seismic gap. The coexistence of this inclined structure and the dominant N-S- to NNW-striking Xiaojiang Fault [54,55] indicates that the fault system in this region is not a simple single-strand strike-slip system but may exhibit compound kinematics, with local zones of reverse-slip accommodation. The NW-trending seismic band (at ~103°00'E in Fig. 1A) strikes at a low angle to the regional maximum principal compressive stress ($\sigma_1$), oriented NW–SE as constrained by regional focal mechanism inversions [56] and broadly consistent with the SSE-directed GNSS velocity vectors in Fig. 1B. Under this stress configuration, the N-S- to NNW-striking Xiaojiang Fault is optimally oriented for left-lateral strike-slip faulting, while the oblique angle between the SSE-directed block motion and the fault strike introduces a fault-normal compressive component, producing local zones of reverse-slip accommodation[57,58]. The cross-sectional profiles, particularly E–E' and F–F', reveal a distinctly inclined seismogenic structure at depths of approximately 10–16 km, with an estimated dip angle of 50°–60°.

This geometry is unlikely to represent a purely vertical strike-slip fault; instead, it suggests the involvement of reverse-slip or oblique-slip components, indicating more complex subsurface kinematics than previously recognized. The coexistence of steeply dipping seismic clusters and a persistent along-strike seismic alignment implies that the fault system may accommodate both lateral and vertical motion.

The Xiaojiang Fault Zone constitutes the eastern boundary of the Sichuan-Yunnan Rhombic Block (SYRB), accommodating the block's southeastward escape and clockwise rotation around the Eastern Himalayan Syntaxis. The "dipping structure" characterized by a dip angle of 50°–60° and merging with the main fault at depth, is kinematically consistent with a positive flower structure or a deep-seated sidewall ramp associated with the Qiaojia pull-apart basin. The literature on adjacent fault systems documents similar "flower structure" geometries in which oblique shortening is accommodated by dipping strands that root into a vertical master fault. The existence of such a structure beneath Qiaojia suggests that the fault system is not merely translating horizontally but is also absorbing significant shortening or extension orthogonal to the fault trace. This geometry creates a mechanism for vertical stress transfer, potentially loading the "compact asperity" identified at 12–18 km depth (Fig. 4B, E). Based on the seismicity pattern — concentrated microseismicity surrounding a relatively quiescent interior — this zone is consistent with the classical 'asperity model' of fault rupture, in which mechanically strong patches remain locked during the interseismic period while surrounding weaker zones deform seismically. Such locked patches often correspond to competent, high-velocity blocks ('knockers') embedded within a generally weaker matrix[59]. However, confirming whether this asperity is indeed associated with a velocity anomaly requires dedicated high-resolution local tomography (Fig.S4). These high-friction patches remain locked during the interseismic period, while surrounding low-velocity zones creep seismically.

Additionally, the fault planes exhibit a distinct bending pattern, indicating that the fault geometry in the deeper section is characterized by curvature or multi-segment connections. Along the profile crossing the reservoir, the most densely distributed seismic events occur in profile A-B (Fig. 4B), where the seismicity exhibits an arcuate, ring-type structure. The source depths are mainly concentrated in the 10–20 km range. The overall morphology suggests the presence of a microseismicity zone at the center, which may be associated with a stress barrier or a small-scale tectonic high. Although the seismicity is dense, no clear planar fault features are observed, indicating that stress is redistributed in the local area, accompanied by complex small-scale faulting processes. The C–C' profile (Fig. 4C) reveals a shallow, nearly planar seismic distribution extending between 15–20 km depth, suggesting a relatively uniform mid-crustal deformation zone without obvious rupture segmentation[60]. A thin, linear distribution of events at shallow depth (~5–10 km) in profile D-D' suggests a sub-horizontal fault or a low-angle deformation layer, possibly acting as a mechanical detachment zone[61]. Since detachment surfaces typically develop at the base of the sedimentary cover and are often associated with reverse-slip activity, the presence of this structure suggests that, in addition to regional strike-slip motion, there may be a significant reverse-slip component or horizontal shear formation.

Profiles D-D', G-G', and H-H' indicate an apparent linear or belt-like pattern in the spatial distribution of seismic events. Such linear distributions are often indicative of the presence of blind faults[13,16,31]. Blind faults, which remain in a low-activity state for long periods, are challenging to detect directly through surface deformation. However, spatial clustering of seismic activity can reveal its existence. Previous studies[31,62] have utilized high-precision seismic catalogs to identify fault activity paths, offering valuable insights into analyzing the tectonic stress field.

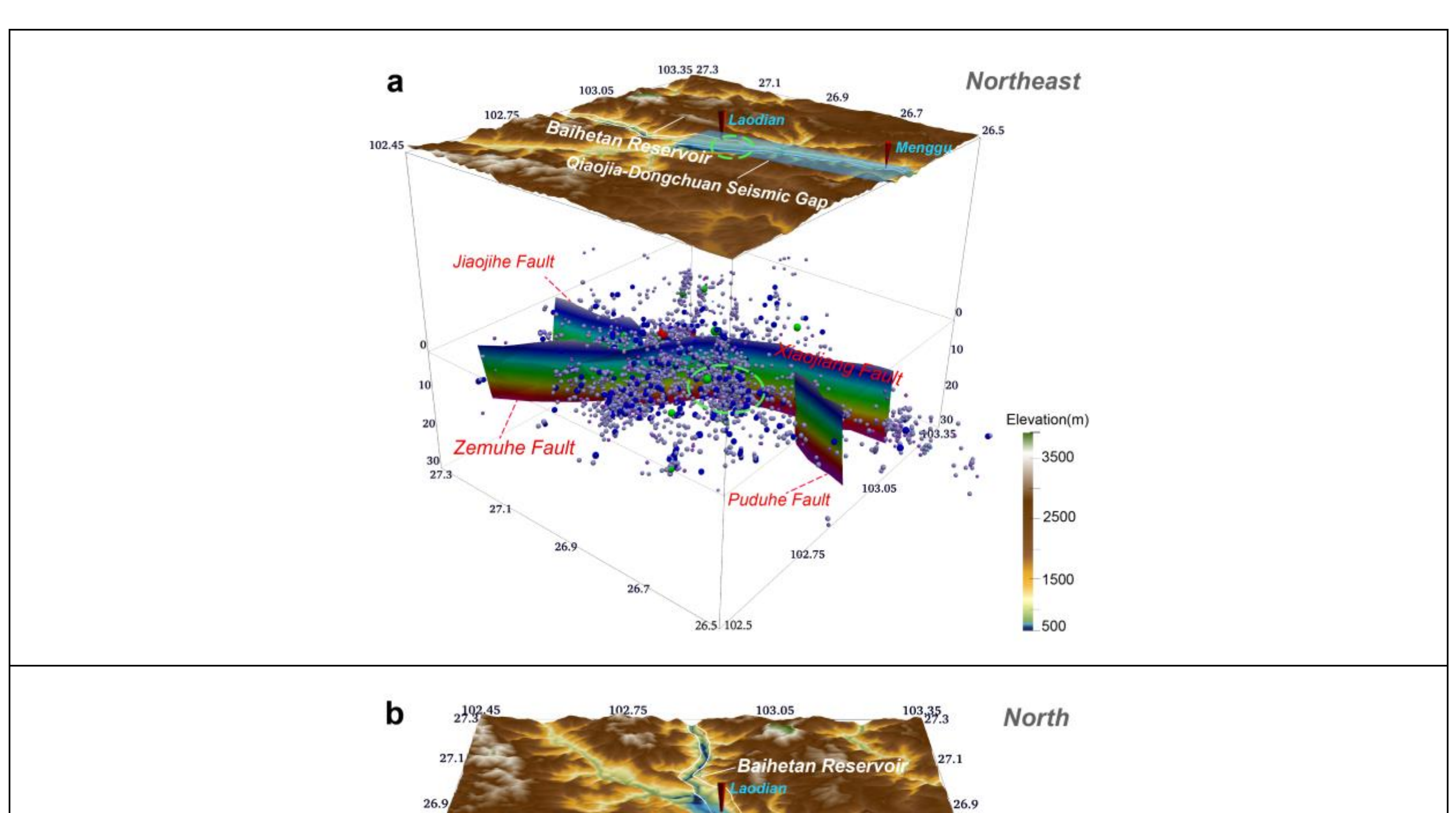
a
Northeast
Laodian
Menggu
Baihetan Reservoir
Qiaojia-Dongchuan Seismic Gap
Jiaojihe Fault
Xiaojiang Fault
Zemuhe Fault
Puduhe Fault
Elevation(m)
3500
2500
1500
500

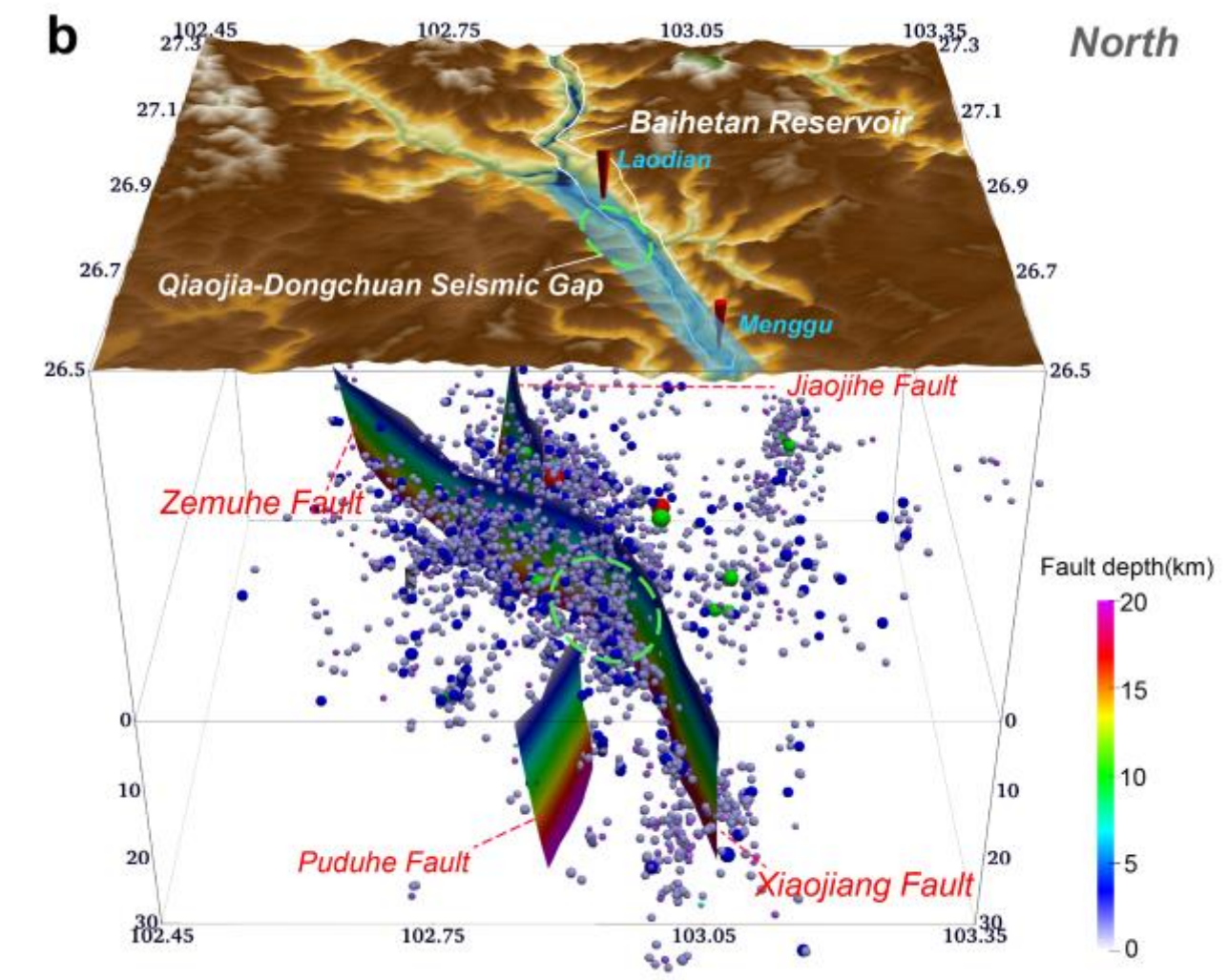
b
North
Baihetan Reservoir
Laodian
Qiaojia-Dongchuan Seismic Gap
Menggu
Jiaojihe Fault
Zemuhe Fault
Puduhe Fault
Xiaojiang Fault
Fault depth(km)
20
15
10
5
0

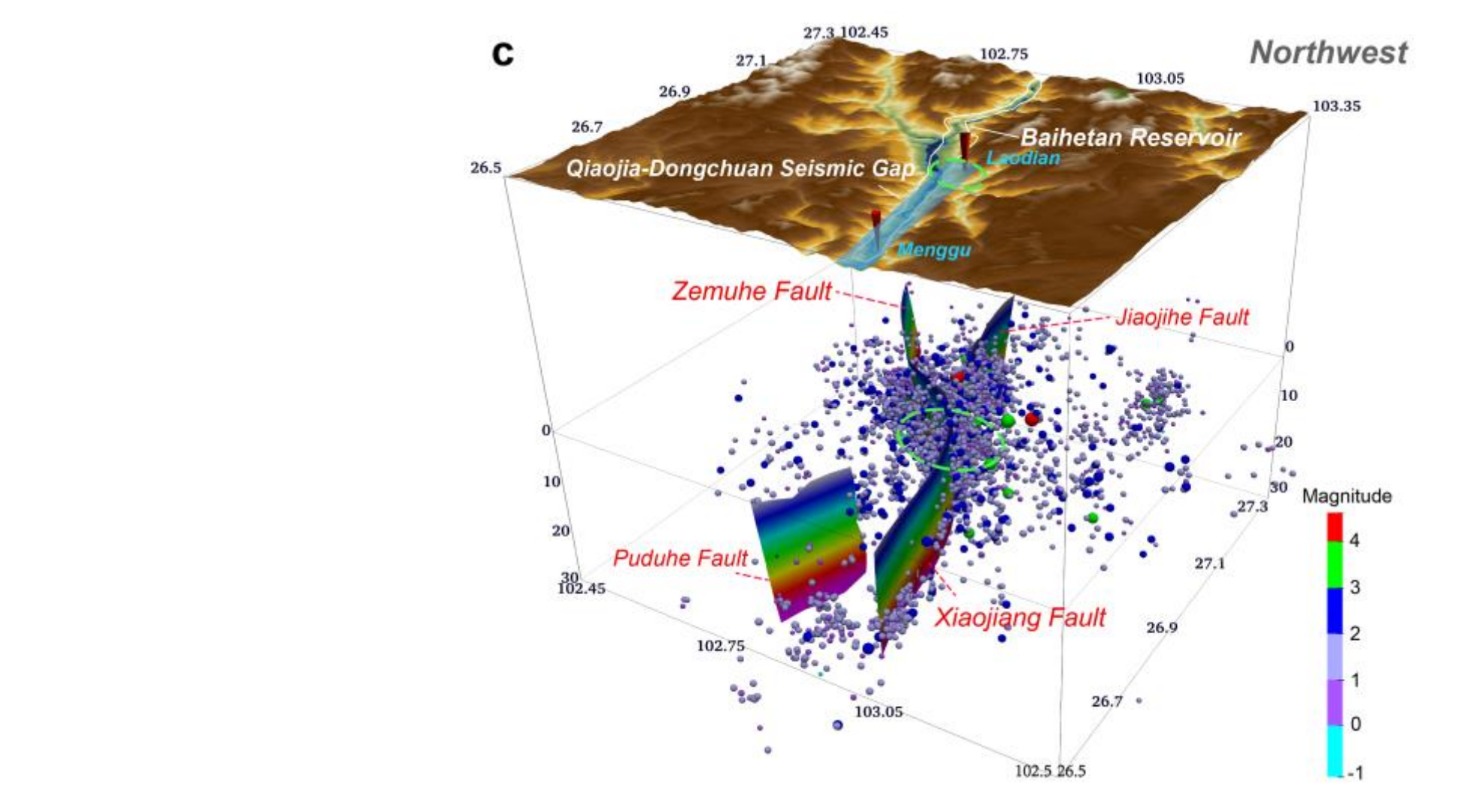
c
Northwest
Baihetan Reservoir
Qiaojia-Dongchuan Seismic Gap
Laodian
Menggu
Zemuhe Fault
Jiaojihe Fault
Puduhe Fault
Xiaojiang Fault
Magnitude
4
3
2
1
0
-1

**Fig. 6.** Three-dimensional views of the topography and seismotectonic architecture in the Qiaojia–Dongchuan region. Panels (**A–C**) show perspective views from the northeast, north, and northwest, respectively. The upper surface displays shaded relief topography. The white line outlines the Baihetan Reservoir, whereas the blue strip marks the Qiaojia–Dongchuan Seismic Gap. The conical symbols indicate the location of the Menggu wharf and Laodian wharf separately. Earthquake hypocenters are shown as spheres, scaled and color-coded by magnitude. The elliptical area highlights a cluster of shallow seismicity, interpreted as a stress concentration or asperity.

As shown in Fig. 6, three-dimensional perspectives (northeast, north, northwest) reveal the topographic and seismotectonic features surrounding the Qiaojia-Dongchuan seismic gap and adjacent regions. Highlighting the three-dimensional spatial distribution of seismic events complements these perspectives. The events exhibit distinct clustering alignments, particularly near the reservoir and major faults, and fault-aligned distribution indicating localized stress concentrations and frequent rock fracturing. Seismicity is predominantly shallow but extends along fault planes, demonstrating depth-dependent fault activity. The spatial correlation between fault traces and event concentrations in 3D confirms the seismic role of these structures (Fig. S4). For instance, dense event zones in Fig. 4E (e.g., profiles E-E') align with fault intersections or specific structural configurations in 3D, supporting the inferences regarding asperities. Notably, the junctions of the Jiaojihe segment along the Daliangshan Fault and Zemuhe Fault exhibit intensified seismicity, likely reflecting complex stress interactions that are conducive to asperity formation. Additionally, the reservoir location coincides with localized event clusters, suggesting that stress perturbations from impoundment may generate secondary asperities.

The presence of a geometrically complex dipping structure has profound implications for rupture dynamics. Historical precedents, such as the 1733 M7.8 Dongchuan earthquake, indicate that the XJF is capable of multi-segment ruptures[5]. Geometric complexities, such as dipping ramps or fault bends, often act as "geometric barriers" that can arrest rupture propagation, thereby limiting earthquake magnitude. Conversely, if the vertical main strand mechanically coupled the dipping structure, a rupture could cascade across both planes, increasing the total slip area and seismic moment release. Evidence from the 2022 Luding earthquake (on the adjacent Xianshuihe fault) showed that the similar geometric asperities[16,63] strongly influenced the rupture propagation. Therefore, evaluating the potential of the dipping structure to act as a rupture gateway or barrier is essential for a comprehensive hazard assessment.

**The Potential of the Strong Earthquake in the Qiaojia–Dongchuan Seismic Gap**

The zones of high-stress regions on active fault planes that accrete significant slip are known as asperities[64] and typically define low $b$-value regimes[65]. In this study, the identification of a localized asperity at 12–18 km depth is particularly significant (Fig. 4B, E), which indicates a mechanically strong domain capable of concentrating tectonic stress. Importantly, this asperity lies immediately adjacent to the Qiaojia−Dongchuan segment, characterized as a seismic gap. There is substantial geodetic consensus supporting the high-locking hypothesis for this segment. Inversions of Global Navigation Satellite System (GNSS) velocity fields consistently show that the Qiaojia-Dongchuan segment exhibits a high coupling fraction (0.8-1.0) (Fig. 5C) down to depths of 15–25 km. These models identify a distinct deficit in slip rate compared to the long-term geological rate (~ 10-13 mm/yr), effectively defining a "locked asperity" capable of generating an Mw 7.2+ event. This configuration—a stress concentration adjacent to a rupture-deficient zone—has been recognized as a characteristic precursor to strong earthquakes in other fault systems worldwide[16,65–67]. If this stress loading is sustained, the seismic gap could represent a potential nucleation point for a future large earthquake.

The Qiaojia–Dongchuan seismic gap has not experienced any moderate-to-large earthquakes in recent decades, despite the high seismic activity of adjacent fault sections. Our Coulomb stress rate calculations indicate that the study area lies within a region of stress loading. The segment exhibits continuous positive Coulomb Failure Stress (CFS) rates along the fault trace, typically ranging from 2 to 4 kPa/yr, with local maxima exceeding 5 kPa/yr near Qiaojia and anomalies exceeding 8 kPa/yr in the surrounding field (Fig. 5A, B). This accumulation rate is physically consistent with geodetic observations. An accumulation of ~10 kPa (0.1 bar) per year implies that, since the last significant event in 1733 (293 years ago), the fault has accumulated approximately 3 MPa (30 bars) of shear stress. This value decreases squarely within the typical stress-drop range (1-10 MPa) for intraplate strike-slip earthquakes, lending quantitative support to the argument that the fault is nearing a critical failure threshold.

A critical observation is that this seismic gap spatially overlaps with the Baihetan Reservoir. However, the lack of seismic clustering around the southern reservoir margin is notable. Instead of the enhanced shallow seismicity typically associated with reservoir-induced activity, the area displays diminished seismic release within the primary fault-controlled zone. This pattern suggests that seismic quiescence is not a consequence of stress relaxation but rather reflects ongoing stress accumulation at depth. Along the western margin of the Xiaojiang Fault, near the reservoir boundary, seismicity exhibits a coherent deep structure. It implies that rupture potential may exceed that inferred from existing moderate-magnitude earthquake sequences.

Decreasing b-values are often interpreted as potential precursors to large earthquakes[68–70]. In our observation window, an increase in shallow seismicity was followed by two M4.0 earthquakes. Subsequently, the b-value dropped below 0.8. Despite a cumulative rise in the number of small earthquakes, the energy released has not counterbalanced ongoing stress accumulation. The Frequency-Magnitude Distribution (FMD) curve (Fig. 3B) further supported this phenomenon. The discrepancy between the theoretical frequency (red line) and the observed catalog (black squares) near the M4.0 range suggests an event deficit. Essentially, earthquakes that "should have happened" based on scaling laws have not occurred, implying they may happen in the future to close the moment budget.

While the "locked asperity" model is compelling, two significant factors complicate the interpretation:

Statistical Limitations: Calculating b-values on short timescales is fraught with statistical peril. Reliable estimation requires $N > 100$ (preferably $N > 500$) to achieve error margins below 0.05. While our total catalog contains >5000 events, the subsets used for monthly time series may drop below 50–100 events. Therefore, observed fluctuations (e.g., the drop to 0.8) could be attributable to stochastic variance rather than physical stress changes.

Creep vs. Locking: The Xianshuihe-Xiaojiang system is known for complex segmentation. Recent InSAR and microseismicity studies suggest parts of the XJF may accommodate deformation through aseismic creep [38,71]. The "dipping structure" identified in this study offers a reconciliation: the main vertical fault trace may be locked (forming the gap), while deep creep occurs along a dipping interface. The clustering of microseismicity at fault junctions could be the manifestation of localized creep transferring stress to these locked patches.

## CONCLUSIONS

Discriminating between the overlapping effects of induced seismicity from anthropogenic activities and natural tectonic activity in areas of a seismic gap — which are inherently prone to hosting strong earthquakes — remains an intriguing and unresolved challenge. This study addresses this challenge through the deployment of a dense array of over 200 seismometers during the peak impoundment phase of the Baihetan Reservoir and the construction of a high-resolution microseismic catalog containing 5,219 events for the Qiaojia–Dongchuan seismic gap — nearly an order of magnitude more than the standard catalog. By integrating this catalog with depth-dependent b-value analysis, Coulomb stress modeling, and cross-sectional structural imaging, we reveal a vertical decoupling mechanism in which shallow seismicity induced by dam impoundment and deep seismicity outlining the locked asperity coexist within the same fault system, yielding three principal findings.

First, we demonstrate that shallow induced seismicity can effectively mask the silent accumulation of deep tectonic strain on the main faults of the seismic gap. The seismicity ensemble exhibits a continuous transition with depth: the shallow portion is characterized by high b-values (>1.0), swarm-like behavior, and strong temporal correlation with reservoir operations — signatures consistent with fluid-driven reservoir-triggered seismicity — while the deep portion (15–20 km) is characterized by low b-values (<0.8) and a continuous seismicity belt outlining a locked asperity on the Xiaojiang Fault, consistent with ongoing tectonic loading. Cross-sectional profiles further reveal a previously unrecognized dipping structure (50°–60°) at 10–16 km depth, indicating compound fault kinematics that may facilitate vertical stress transfer between these two depth regimes.

Second, the computed stress accumulation suggests that this seismic gap is in a critical state with elevated rupture potential. The CFS rate along the Qiaojia–Dongchuan segment reaches 7–10 kPa/yr, implying an accumulated stress of approximately 3 MPa since the last significant earthquake in 1733 — a value within the typical stress-drop range (1–10 MPa) for intraplate strike-slip earthquakes. The frequency-magnitude distribution further reveals a deficit of M ≥ 4.0 events relative to the statistical expectation from the Gutenberg–Richter scaling, suggesting that the moment budget has not been closed by the current seismicity.

Third, this decoupling model provides a new conceptual framework for assessing seismic risks in reservoir-fault coupled systems. The recognition that both mechanisms coexist — with their relative dominance varying systematically with depth — carries a practical implication: elevated shallow seismicity following impoundment should not be interpreted solely as local stress release but must be evaluated alongside indicators of deep tectonic strain accumulation to avoid underestimating the hazard posed by a locked fault at depth.

## Methods

### AI-Powered Earthquake Detection and b-value

Waveform data, spanning 200 continuous days from August 25, 2022, to March 13, 2023, were collected from the Qiaojia-Dongchuan region of Yunnan Province (Fig. 2B). This dataset was acquired using 10 broadband seismometers and 201 short-period seismometers, strategically deployed with a minimum interstation distance of approximately 5 km. While most continuous data were collected using broadband instruments capable of capturing long-period signals due to their wide frequency response, short-period seismometers also contributed. Each station recorded three-component waveforms (N, E, Z) at a sampling rate of 100 Hz.

Deep learning (DL) has emerged as a standard practice in modern seismology, demonstrating significant success in applications such as earthquake detection and phase-picking for routine seismicity monitoring[72,73]. Here, we compiled a DL workflow for processing raw data and creating a high-resolution seismicity catalog. To maximize detection accuracy, we implemented an ensemble workflow that integrates three complementary models — PhaseNet[74], EQTransformer[73], and TranSeis[75]. PhaseNet was trained on data from the Northern California Earthquake Data Center, and EQTransformer was trained on the global STEAD dataset[76]. Studies have shown that transfer learning or region-specific training is crucial for maintaining high precision and recall in the complex, scattering-rich crust of the Tibetan Plateau[77,78]. It is where TranSeis, which is explicitly trained on Chinese seismic datasets[79], can help improve performance on mainland Chinese data. All models were applied to recorded continuous data separately, and duplicate detections were removed to produce a consolidated, high-confidence seismic catalog. The ensemble approach mitigates the weaknesses of individual models. PhaseNet typically offers high recall (detecting many events) but can be prone to false positives in high-noise environments. EQTransformer generally provides higher precision but may miss events with a lower signal-to-noise ratio (SNR) (Table S1). By combining these with TranSeis, we achieved a significantly more complete catalog (lower Magnitude of Completeness, Mc) than would be possible with any single one. The ensemble approach significantly enhances model generalization and prediction accuracy on cross-domain data, achieving performance comparable to that of transfer learning without requiring new labeled datasets or retraining. Additionally, it suppresses noise and reduces picking errors—by nearly 50% in some cases—thereby minimizing false positives compared to individual base models[80].

Subsequently, we use the Bayesian Gaussian Mixture Model association method, GAMMA[81], for event association and Hypoinverse[82] to obtain absolute locations. HypoDD[83] was then used to enhance the hypocentral location accuracy through cross-correlation and time-residual relative relocation procedure. The velocity model used for earthquake localization is a newly updated 3-D reference P- and S-wave community velocity model of the crust and uppermost mantle in southwest China[84].

The magnitude of detected seismic events is estimated using the new local magnitude formula proposed by Yang *et al.*[85], which is suitable for calculating the magnitude of small earthquakes from dense-array data and ideal for the Sichuan Basin.

$$M_L = log_{10}(A) + 1.26 log_{10}(D) - 0.0026D - 2.2, \quad (1)$$

where $A$ is the amplitude of the waveform, $D$ is the hypocentral distance.

**Frequency-Magnitude Analysis**

The b-value is a key parameter that can be computed from the Frequency-Magnitude Distribution (FMD) law introduced by Gutenberg & Richter[36]. The FMD law is expressed as:

$$log_{10}N = a - bM, \quad (2)$$

where $N$ denotes the cumulative number of earthquakes having magnitudes equal to and larger than $M$, while $a$ and $b$ are constants that may vary across space and time.

We applied the Maximum Curvature (MAXC) method[86], which defines $M_c$ as the magnitude of the highest frequency in the non-cumulative frequency-magnitude distribution. We use a moving window approach with a constant sample size of 500 events and a magnitude bin width of 0.1. To quantify the uncertainty of $M_c$, we utilized a bootstrap approach with 100 iterations, calculating the standard deviation of the computed $M_c$ values for each grid point[87].

The constant $a$ reflects the level of seismic activity or earthquake productivity, whereas the b-value (the slope of the log-linear relation) indicates the relative distribution of earthquake magnitudes. A decrease in b-value generally implies a reduction in small-magnitude events, suggesting increasing stress accumulation. In contrast, a higher b-value may reflect enhanced seismic activity and possible stress release[65]. The b-value is used as a proxy for stress measurement to analyze the spatio-temporal characteristics of seismicity in region[88]. It helps assess whether stress is being released gradually through small quakes or accumulating for a larger event. In this study, we calculated the $M_c$, the b-value, and their temporal evolutions using ZMAP[37], a widely used and reliable tool for b-value estimation. Detailed ZMAP parameter settings are listed in Table S2 of the Supplementary Information.

**Coulomb Failure Stress Accumulation Rate and Triggering Probability Modeling**

To further assess the state of stress accumulation and seismic hazard along the Qiaojia–Dongchuan segment of the Xiaojiang Fault Zone, we constructed a regional map of the annual Coulomb failure stress (CFS) accumulation rate using observed GNSS velocities[35]. We applied a layered Maxwell viscoelastic model[89] to compute the full strain components and the corresponding stress-rate tensors. The shear modulus was set to $30 \times 10^9$ Pa and the Poisson's ratio to 0.25[89,90]. At the same time, the viscosity of the lithosphere is based on the results of Shi and Cao[91] (see Table S3 in the Supplementary Information for more details).

$$\Delta \mathrm{CFS} = \Delta\tau + \mu'\left(\Delta\sigma_{\mathrm{n}} + \Delta \mathrm{P}_{\mathrm{p}}\right) \tag{3}$$

Here, $\Delta\tau$ denotes the change in shear stress on the fault plane, and $\mu'$ is the coefficient of friction on the fault plane, assumed as 0.4. $\Delta\sigma_{\mathrm{n}}$ represents the change in normal stress on the fault plane, and $\Delta \mathrm{P}_{\mathrm{p}}$ denotes the change in pore pressure. The pore pressure was estimated using a highly simplified model, in which the influence of water-level variation on pore pressure is assumed to decay exponentially with distance from the well location, Menggu well in our study (Fig. 3A, Fig. 6), with a decay radius of 5 km.

The fault-normal and shear components of the CFS rates were resolved onto the fault-plane geometry of the Qiaojia–Dongchuan segment, referring to a dip angle of 85° and a strike of 167°[92], as constrained by geological and seismotectonic observations. To assess the robustness of the modeled stress–strain field, we reconstructed the regional velocity field from the computed strain rates and compared it with the observed GNSS velocity field. The horizontal displacements along the lateral boundaries were interpolated from the observed horizontal GNSS velocity field. The modeled velocities show strong agreement with the observations in both spatial distribution and orientation, particularly across regions of pronounced velocity gradients near the fault (Fig. 7).

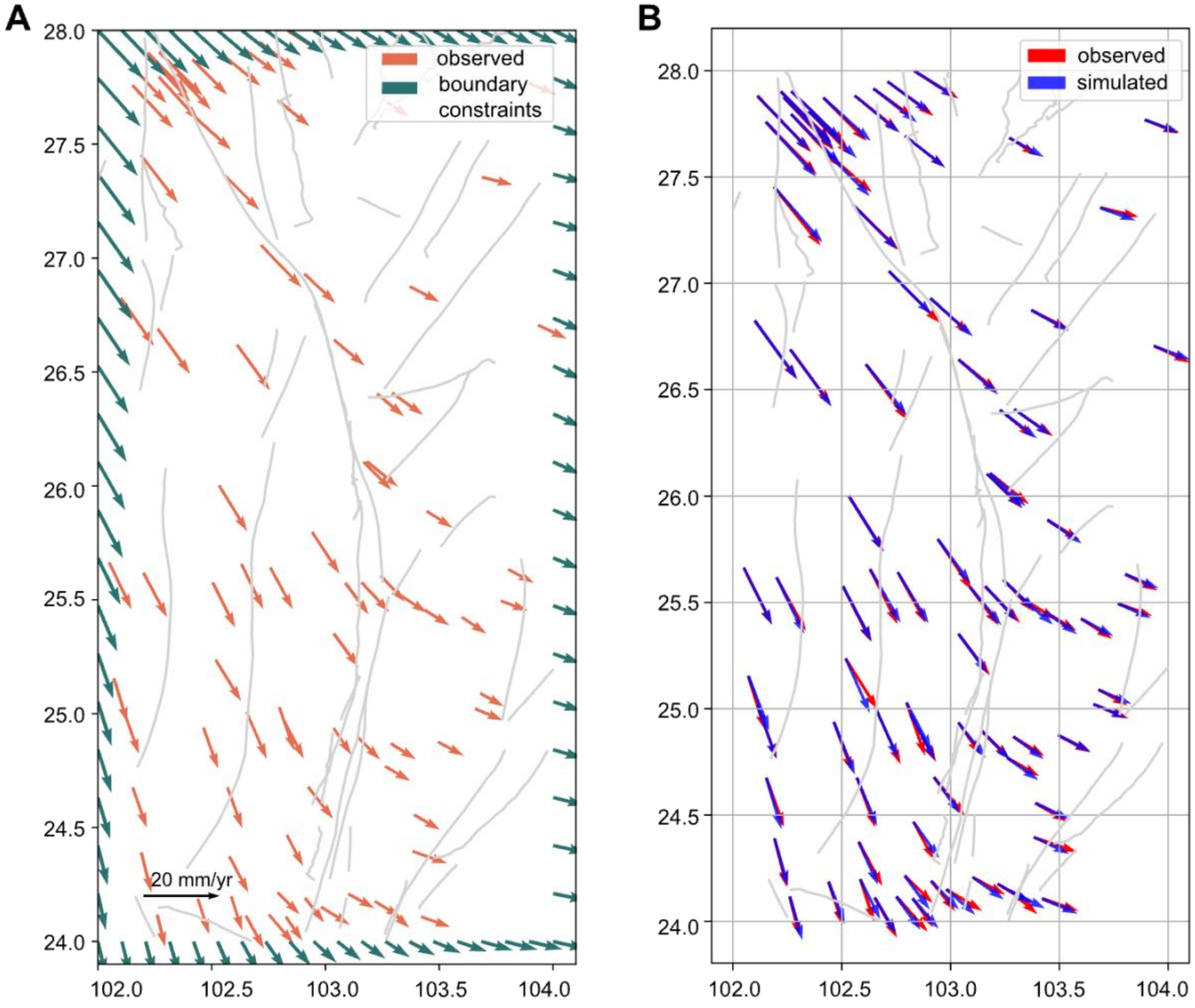

**Fig. 7.** Validation of simulated GNSS velocity field and boundary constraints. (**A**) Observed GNSS velocity vectors (red) and boundary constraint vectors (green) used for strain-driven velocity field reconstruction. (**B**) Comparison between observed (red) and simulated (blue) GNSS velocities across the study area.

Furthermore, using a rate-and-state friction model[23,93,94], we estimated earthquake-triggering probabilities across the region. This model, developed from frictional constitutive equations, describes the influence of stress perturbations on regional seismic activity and evaluates the roles of the associated parameters, thereby enabling the calculation of earthquake occurrence rates.

$$R(t) = \frac{\gamma}{\exp\left(-\frac{\Delta CFS}{A_\sigma} - 1\right) \cdot \exp(-\frac{t}{t_\varepsilon}) + 1},$$

$$\mathrm{P} = 1 - \exp\big(-\mathrm{R}(t)\big) \tag{4}$$

In this study, to investigate long-term stress accumulation, the time interval $t_\varepsilon$ was set to 6 years, the direct effect parameter $A_\sigma$ is typically determined experimentally, commonly ranging between 0.0012 and 0.6 MPa[95], and it was set to 0.3 MPa, the characteristic relaxation time $t_\varepsilon$ to 10 years, and the background seismicity rate constant $\gamma$ to 0.01.

## Data availability

The software and original waveforms associated with this manuscript are licensed under MIT and published on Zenodo https://zenodo.org/records/15629731. The most recent fault data are available at http://www.cses.ac.cn/sjcp/ggmx/2024/609.shtml, and the latest velocity model used is the High Accuracy Velocity Model version 2.0, available at http://cses.ac.cn/sjcp/ggmx/2022/589.shtml. The GNSS velocity and strain rate data are available at https://zenodo.org/records/10215151.

## Acknowledgments

We are grateful to the two anonymous reviewers for their insightful feedback and helpful suggestions. This work is supported by the National Natural Science Foundation of China (No. U2239205). S. M. M is supported by the Harvard Milton Fund. We thank Dr. Zhang Zhengfeng for kindly providing the GNSS velocity and strain rate data used in this study. We are grateful to the ZMAP team for developing the ZMAP tool.

## Author contributions

Conceptualization: Y. Zhou, H. Zhang and S. M. Mousavi. Methodology: Y. Zhou, H. Zhang and S. M. Mousavi. Validation: Y. Zhou, S. M. Mousavi and H. Zhang. Formal analysis: Y. Zhou. Investigation: Y. Zhou. Software: Y. Zhou. Visualization: Y. Zhou, P. He. Resources: H. Zhang, Y. Shi. Data curation: Y. Zhou. Writing—original draft: Y. Zhou. Writing—review and editing: Y. Zhou, H. Zhang, S.M. Mousavi and G. Yin. Supervision: H. Zhang and S.M. Mousavi. Project administration: H. Zhang, G. Yin, Y. Guo and S. Yi. Funding acquisition: H. Zhang, Y. Shi.

## Competing interests

The authors declare that they have no competing interests.